\documentclass{sigchi-ext}
\usepackage[T1]{fontenc}
\usepackage{textcomp}
\usepackage[scaled=.92]{helvet} 
\usepackage{graphicx} 
\usepackage{balance}  
\usepackage{booktabs} 
\usepackage{ccicons}  
\usepackage{ragged2e} 
\usepackage{enumitem}

\def\plaintitle{SIGCHI Extended Abstracts Sample File: Note Initial
  Caps} 
\def\emptyauthor{}
\def\plainkeywords{Explainable AI; Human-Centered AI; Cybersecurity; Domain-Specific Workflows; Visualization.}

\title{More Questions than Answers? Lessons from Integrating Explainable AI into a Cyber-AI Tool}

\numberofauthors{6}
\author{%
  \alignauthor{%
    \textbf{Ashley Suh}\\
    \affaddr{MIT Lincoln Laboratory} \\
    \affaddr{Lexington, MA, USA} \\
    \email{Ashley.Suh@ll.mit.edu} }\alignauthor{%
    \textbf{Kenneth Alperin}\\
    \affaddr{MIT Lincoln Laboratory} \\
    \affaddr{Lexington, MA, USA} \\
    \email{Kenneth.Alperin@ll.mit.edu}} \vfil \alignauthor{%
    \textbf{Harry Li}\\
    \affaddr{MIT Lincoln Laboratory} \\
    \affaddr{Lexington, MA, USA} \\
    \email{Harry.Li@ll.mit.edu} }\alignauthor{%
    \textbf{Steven R.~Gomez}\\
    \affaddr{MIT Lincoln Laboratory} \\
    \affaddr{Lexington, MA, USA} \\
    \email{Steven.Gomez@ll.mit.edu} } \vfil \alignauthor{%
    \textbf{Caitlin Kenney}\\
    \affaddr{MIT Lincoln Laboratory} \\
    \affaddr{Lexington, MA, USA} \\
    \email{Caitlin.Kenney@ll.mit.edu} } }

\definecolor{linkColor}{RGB}{6,125,233}
\hypersetup{%
  pdftitle={\plaintitle},
  pdfauthor={\emptyauthor},
  pdfkeywords={\plainkeywords},
  bookmarksnumbered,
  pdfstartview={FitH},
  colorlinks,
  citecolor=black,
  filecolor=black,
  linkcolor=black,
  urlcolor=linkColor,
  breaklinks=true,
}

\usepackage{hyperref}

\begin{document}

\CopyrightYear{2020}
\setcopyright{rightsretained}
\conferenceinfo{ACM CHI 2024 Workshop on Human-Centered Explainable AI (HCXAI)}{May 12, 2024, Honolulu, HI, USA}
\isbn{978-1-4503-6819-3/20/04}
\doi{https://doi.org/10.1145/3334480.XXXXXXX}
\copyrightinfo{\acmcopyright}

\maketitle

\RaggedRight{} 

\begin{abstract}
  We share observations and challenges from an ongoing effort to implement Explainable AI (XAI) in a domain-specific workflow for cybersecurity analysts. Specifically, we briefly describe a preliminary case study on the use of XAI for source code classification, where accurate assessment and timeliness are paramount. We find that the outputs of state-of-the-art saliency explanation techniques (e.g., SHAP or LIME) are lost in translation when interpreted by people with little AI expertise, despite these techniques being marketed for non-technical users. Moreover, we find that popular XAI techniques offer fewer insights for real-time human-AI workflows when they are post hoc and too localized in their explanations. Instead, we observe that cyber analysts need higher-level, easy-to-digest explanations that can offer as little disruption as possible to their workflows. We outline unaddressed gaps in practical and effective XAI, then touch on how emerging technologies like Large Language Models (LLMs) could mitigate these existing obstacles.
\end{abstract}




\section{Motivation}

While AI has been applied successfully in several tasks related to cybersecurity (e.g., flagging emails as potential spam~\cite{xaiPhishing2023}), its use in tasks that are primarily driven by human operators requires a sufficient degree of interpretability before it can be adopted for real-world use~\cite{Human_AI_Partnership_2023}. Srivastava et al.~provide an overview of some of these challenges for AI adoption in the cyber domain~\cite{srivastava2022xai}.

For cyber operations specifically, the use of Explainable AI (XAI) presents a unique challenge, as the end users of AI decision-support tools are not necessarily AI experts; they tend to be highly skeptical about AI altogether~\cite{Ulery_2020}; and the system behaviors they analyze (e.g., network or software behavior) are highly context dependent.

As a result, a motivating question for the Human-Centered XAI community is: \textit{Do current XAI techniques effectively support users in cyber-related analysis tasks, and are there significant remaining challenges that need to be addressed?} We reflect on these questions following a preliminary study into improving interpretability for a source code classifier using model-agnostic, local explainers.

\begin{marginfigure}[-5pc]
  \begin{minipage}{\marginparwidth}
    \centering
    \includegraphics[width=.95\columnwidth]{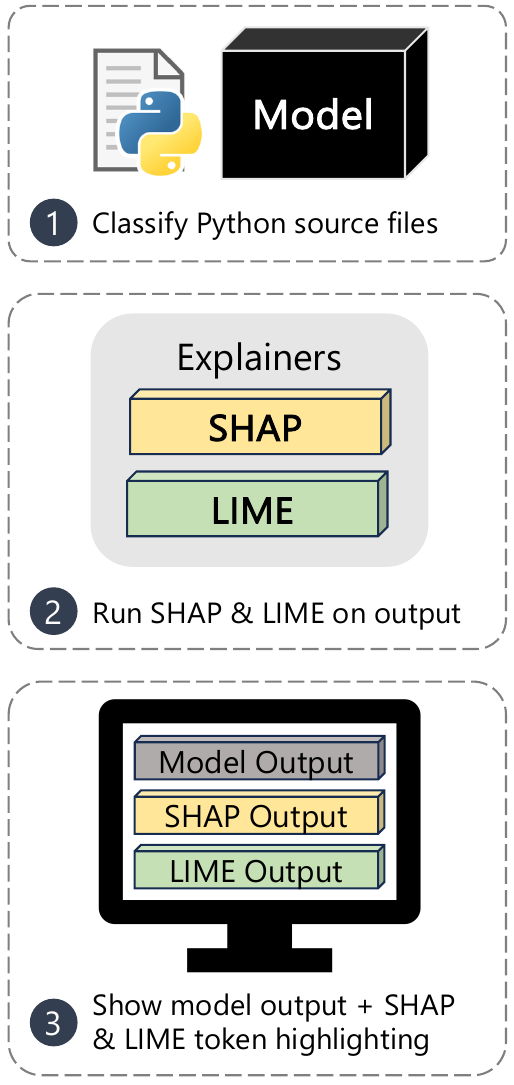}
    \caption{System architecture: the model classifies Python source code files as implementing ML or not. We use XAI to provide tokenized highlighting to cyber analysts who are sanity checking a file and its classification.}~\label{fig:architecture}
  \end{minipage}
\end{marginfigure}

\section{Case Study: XAI for a Source Code Classifier}

\begin{figure*}[t]
  \centering
    \includegraphics[width=0.8\textwidth]{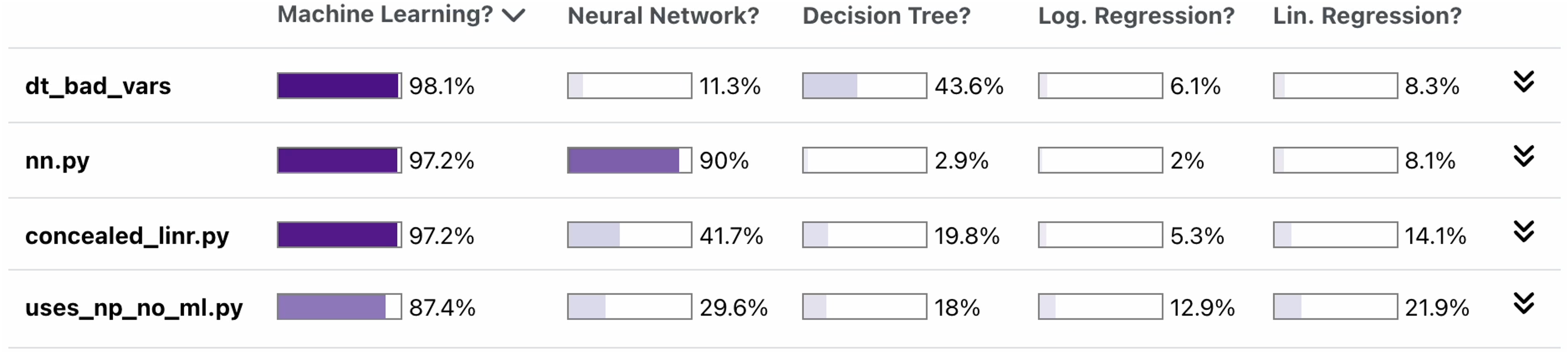}
  \caption{Before presenting any saliency explanations, a table presents the classification confidence scores indicating how likely the source code file implements ML, or a subtype of ML.}~\label{fig:dashboard}
\end{figure*}

We are broadly interested in understanding how to design effective AI-driven decision support for cyber analysts whose jobs are to test software systems, identify issues, and decide on appropriate courses of action. Here, our goal is to understand whether off-the-shelf, widely-used XAI techniques provide useful transparency to a neural network classifier (the ``model"). While we did not create this model, we were tasked to illuminate how XAI can (or cannot) support explaining this model's classification of source code artifacts. The model was trained to read a collection of Python source code files and classify them as either ``\textbf{Yes}, this Python file has code that implements ML" or ``\textbf{No}, this Python file does not have code that implements ML." The purpose of the model was to assist analysts in identifying structures in the source files that may need additional testing. It also predicts the presence of ML sub-types in the implementation. Figure~\ref{fig:dashboard} shows an example of how these basic classifications are displayed.

We integrated SHAP~\cite{lundberg2017unified} and LIME~\cite{ribeiro2016should} into a decision-support web application to explain the model behavior on instances of source code. In particular, we used SHAP and LIME to compute values for highlighting within a Python file, indicating whether a particular word or set of words from the file contributed to or contradicted the model's prediction. 

\begin{figure}
  \centering
  \includegraphics[width=\columnwidth]{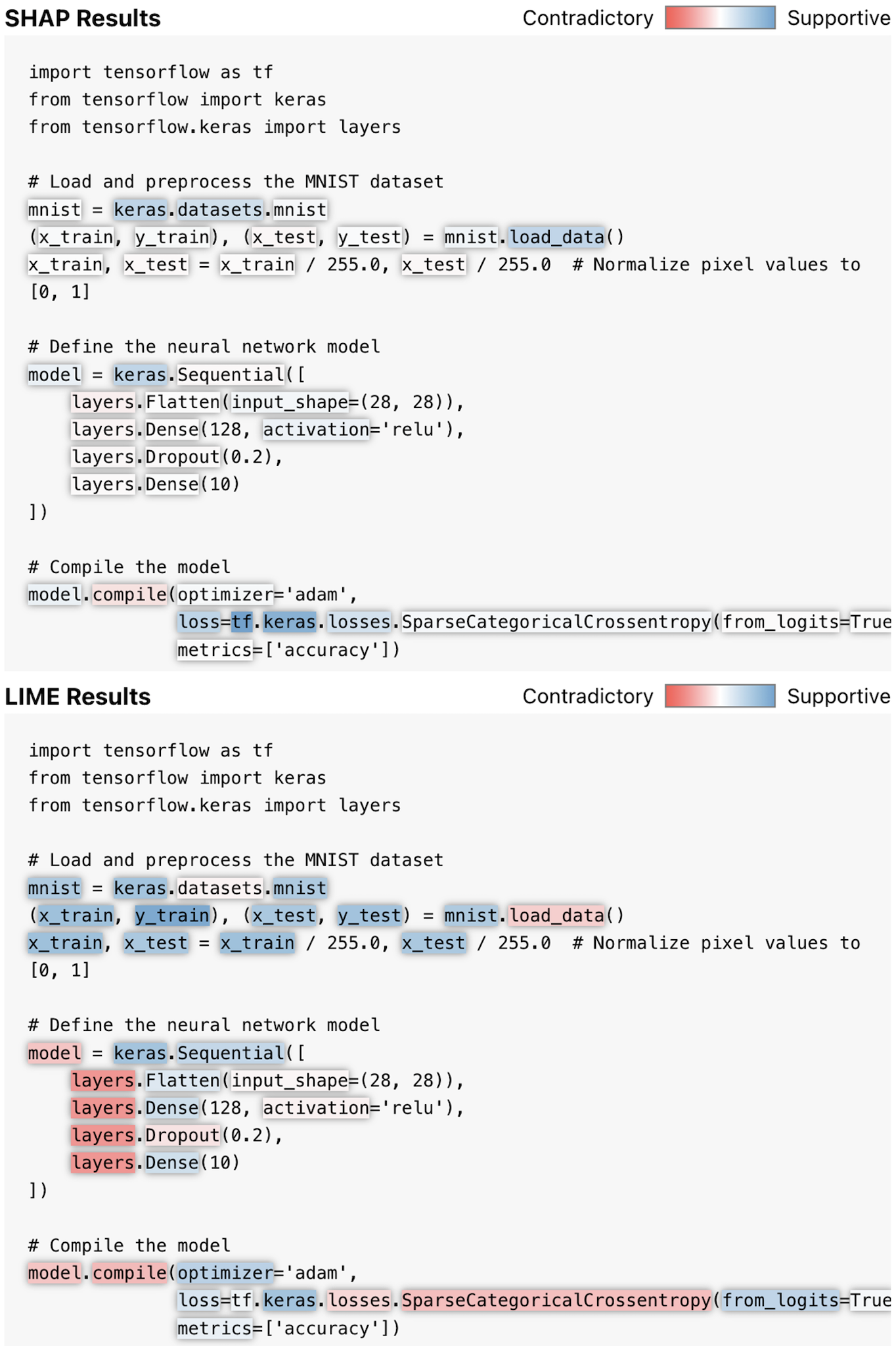}
  \caption{Visual highlighting for a Python file based on SHAP and LIME explanations of the ``implements ML?" decision.}~\label{fig:shap}
\end{figure}

SHAP and LIME are widely marketed as interpretable solutions to explain black box models~\cite{keita2021xai}, in particular to discover comparable local feature-importance values (e.g., saliency) for a black-box model. We used both explainers to provide some redundancy, and to investigate how agreement and conflict between the methods are understood by users. 
Moreover, we hoped that by situating the saliency values into the domain itself (i.e. visually highlighting discriminatory text in source code files) then end-users would have a better understanding of the explanations. We found overall that end-users struggled to comprehend the explanations provided by SHAP and LIME, we discuss this challenge later. Consequently, integrating additional explainers (other gradient methods~\cite{nadeem2023sok}, counterfactual explanations~\cite{wachter2017counterfactual}, etc.) that have shown promise for related applications is an essential next step. The architecture for our entire pipeline is shown in Figure~\ref{fig:architecture}, and an example visualizing salient keywords is shown in Figure~\ref{fig:shap}.

While we have not yet performed a formal evaluation, below we share our early impressions based on iterative demonstratives to our stakeholders and others who are representative of our target analyst end users (henceforth, ``users").

\section{Positive Observations}
\textbf{Explainability is necessary for cyber.} We find that both stakeholders and users consider it helpful to have \textit{any kind} of explanation provided for a model's outputs, increasing confidence in the use of AI altogether. 

\textbf{Explanations help analysts understand a model's behavior.} We find that users were capable of pointing out interesting behaviors of the classifier based on the outputs of SHAP and LIME. For example, one user asked us, ``\textit{should the model really be using words like} \texttt{def} \textit{or} \texttt{print} \textit{for its prediction?}''

\textbf{XAI bridges gaps in AI expertise, creating a dialogue.} As SHAP and LIME provide users some transparency in what keywords the classifier is learning for its decision making (e.g., the use of \texttt{numpy} for classifying a file as implementing ML), added explainability can open up a dialogue for non-AI experts (e.g., cyber analysts) and the developers of these models during collaborative workflows.

\section{Top Challenges}
\textbf{Explainers contradict each other, promoting distrust and confusion by users.} Indisputably, the biggest challenge we ran into was the disagreements between explainers. In some cases, SHAP would highlight particular tokens as strongly supportive, while LIME would highlight those same tokens as highly contradictory. An example of these disagreements can be seen in Figure~\ref{fig:shap}. We had also hoped that combining XAI techniques would lead to more robust explanations~\cite{xaiForEndUsers2023}---a known challenge~\cite{alvarez2018robustness}---but the disagreements between techniques left users confused. We posit that these disagreements between XAI techniques can promote distrust in the model's outputs. While a level of distrust or skepticism is necessary when interpreting a model's predictions for cyber defense (after all, a model will not always be correct) this does not seem like a desirable effect of using an ensemble of XAI techniques. 

\textbf{XAI, on a conceptual level, is still confusing to users and experts alike. } Analysts and developers expressed confusion between the distinction of the model itself and the explanation outputs. After clarifying this to cyber analysts, they tended to expect that the outputs of SHAP and LIME would be responsible for `retraining' the model and correcting the classifier's mistakes, without human intervention.

\textbf{Off-the-shelf XAI options are insufficient for cyber.} We find that the post-hoc and localized nature of XAI techniques for black-box models leave users dissatisfied. They stressed the need to understand a model's outputs in-situ, and on data the model had not been seen before. Moreover, users expressed the need for higher-level interpretations of the explanations, as to not interrupt their workflows. 

\textbf{Higher-level visual abstractions are needed for better end-user interpretability.} We found that manually inspecting the highlighted tokens provided by SHAP and LIME for each source code file is a burdensome task for analysts. Importantly, the typical visualizations provided by explainers require a high level of visual literacy, insufficient for our user group. Future work must address how higher-level visual abstractions -- perhaps an ensemble-based approach to illustrate multiple factors or features -- can be used to reduce time spent interpreting explanations.

 
\section{Broader Reflections for Human-Centered XAI}
\label{sec:needs}

\textbf{We ought to be more transparent in the \textit{actual} expertise required for interpreting explainers.} 
Common XAI techniques like SHAP and LIME are consistently cited as methods that ``\textit{promote trust and understanding}'' for stakeholders, decision-makers, and ``\textit{non-technical}'' end-users (e.g.,~\cite{keita2021xai}). However, we observed (as have others in the HCXAI community) that these techniques are insufficient for actual ``non-technical'' end-users.

Our community should work towards understanding why this misconception continues to be perpetuated, and how we can mitigate it. Is it because techniques like SHAP and LIME are the most widely available to model developers? Is it because they are \textit{actually} the best we can do to explain black box models, so developers are forced to market SHAP and LIME as ``interpretable solutions'' in order to gain model acceptance from stakeholders?

If we want XAI to be widely accessible beyond the expert data science community, two things are imperative. First, we must contribute XAI techniques that can address meaningful end-user questions -- \textit{not} simple questions like ``what kind of data are you modeling?'' and ``do you want local or global explanations?'' We provide example questions in our sidebar. Second, contributed XAI techniques ought to incorporate potentially \textit{many} visualizations that are both communicative and easy to digest -- 
a single visualization to illustrate an explainer (e.g., a SHAP diagram) is insufficient.

 \marginpar{%
   \vspace{-80pt} \fbox{%
     \begin{minipage}{0.9\marginparwidth}
     \centering
     \vspace{0.2cm}
        \small
            \centering 
            \textbf{General Workflow Questions} \\
            \begin{itemize}[leftmargin=*,topsep=2pt, partopsep=0pt,itemsep=2pt,parsep=2pt]
                \item How much time does a user have to interpret the XAI?
                \item How critical is the accuracy of the final decision to the broader system/workflow?
                \item How much can a human interact with the model?
            \end{itemize}
      \vspace{0.4cm}
      \textbf{Questions for Interpreting \\Model Outputs} \\
          \begin{itemize}[leftmargin=*,topsep=2pt, partopsep=0pt,itemsep=2pt,parsep=2pt]
            \item How much uncertainty is there in the models outputs?
            \item When will the model fail? What are its limitations?
            \item What are the shortcomings of its training data?
            \item How much time is there to make a decision?
            \item What level of risk is acceptable for the model's use?
         \end{itemize}
     \vspace{0.2cm}
     \label{sec:sidebar}
 \end{minipage}}
 }
 
\textbf{When explanations leave users with more questions than answers, can we leverage dialogue systems?}
The current state of commonly-used explainers like SHAP and LIME will likely leave end-users with more questions than answers. This could present an opportunity for us to incorporate additional tools, like conversational agents, that help users interpret the outputs of these XAI methods. Emerging generative Large Language Models and Vision-Language Models (LLMs and VLMs) could soon facilitate question-answering about models \textit{and} explanations. 

Importantly, future research into XAI-focused conversational agents should consider the essential human factors needs of the particular domain the model is deployed for. For example, in a high-stakes, high-risk environment like cybersecurity, ``\textit{the simple classification result is not the essential information that the [operator] requires, instead, they need to understand more about the threat and the reason for it to be treated as [such]}.''~\cite{srivastava2022xai} 

Beyond interpreting explanations, it is also possible that an LLM could participate in a back-and-forth dialogue with an end-user to understand model requirements that will lead to acceptance. For example, an LLM could facilitate questions like ``how much risk are you willing to take on with this model?'' or ``how much time are you willing to spend interpreting the model's outputs?'' Illuminating the answers to these questions can inform downstream modeling and development, and potentially alleviate the burden of this back-and-forth for data scientists.

An essential unknown that will need to be addressed before the incorporation of these tools is how an LLM might output false causalities, correlations, or explanations when answering a user's questions (often referred to as `hallucinations'). For example, if the user asks the LLM to walk through the reasoning process of a model's decision -- a process that is actually unknown -- the LLM may falsify information to satisfy the user, perpetuating a misunderstanding of the model's actual capabilities.

\section{Acknowledgements}
DISTRIBUTION STATEMENT A. Approved for public release. Distribution is unlimited. This material is based upon work supported by the Under Secretary of Defense for Research and Engineering under Air Force Contract No. FA8702-15-D-0001. Any opinions, findings, conclusions or recommendations expressed in this material are those of the author(s) and do not necessarily reflect the views of the Under Secretary of Defense for Research and Engineering. © 2024 Massachusetts Institute of Technology. Delivered to the U.S. Government with Unlimited Rights, as defined in DFARS Part 252.227-7013 or 7014 (Feb 2014). Notwithstanding any copyright notice, U.S. Government rights in this work are defined by DFARS 252.227-7013 or DFARS 252.227-7014 as detailed above. Use of this work other than as specifically authorized by the U.S. Government may violate any copyrights that exist in this work.


\balance{} 

\bibliographystyle{SIGCHI-Reference-Format}
\bibliography{xai}

\end{document}